
\magnification=1200
\def\f#1#2{{\textstyle {#1\over#2}}}
\def\d{\partial}
\def\next{\hfil\break\noindent}
\def\R{{\bf R}}
\font\title=cmbx12
\def\BeweisEnde{\hskip 1mm
  \hbox{\vrule height 1.9mm width 2mm depth 0.1mm}}
\hfuzz=3pt
\font\title=cmr10    scaled \magstep4
\parskip=5pt plus 5pt
\parindent=0cm
\overfullrule=0pt

{\title
\centerline{The cosmic no-hair theorem and the}\vskip 0.3cm
\centerline{  nonlinear stability of homogeneous} \vskip 0.3cm
\centerline{Newtonian cosmological models}
}
\vskip 1cm
\centerline{Uwe Brauer$^*$,
Alan Rendall$^*$ and
Oscar Reula$^{*\dagger}$\footnote{$^{\sharp}$}{Member of CONICET}}
\vskip 0.5cm
 \centerline{$^*$ Max-Planck-Institut f{\"u}r Astrophysik,}
 \centerline{Karl Schwarzschild Str. 1}
 \centerline{  85 740 Garching }
 \centerline{Germany}
\vskip 0.5cm
 \centerline{$^{\dag}$FaMAF, UNC,}
  \centerline{Ciudad Universitaria,}
 \centerline{5000 C\'ordoba}
 \centerline{Argentina}
\vskip1cm
{\bf Abstract}\next
The validity of the cosmic no-hair theorem is investigated in
the context of Newtonian cosmology with a perfect fluid matter
model and a positive cosmological constant. It is shown that
if the initial data for an expanding cosmological model of
this type is subjected to a small perturbation then the
corresponding solution exists globally in the future and
the perturbation decays in a way which can be described
precisely. It is emphasized that no linearization of the
equations or special symmetry assumptions are needed.
The result can also be interpreted as a proof of the nonlinear
stability of the homogeneous models. In order to prove the
theorem we write the general solution as the sum of a
homogeneous background and a perturbation. As a by-product of
the analysis it is found that there is an invariant sense in
which an inhomogeneous model can be regarded as a perturbation
of a unique homogeneous model. A method is given for associating
uniquely to each Newtonian cosmological model with compact spatial
sections a spatially homogeneous model which incorporates its
large-scale dynamics. This procedure appears very natural in
the Newton-Cartan theory which we take as the starting point
for Newtonian cosmology.

\vfill
\eject

\noindent
{\bf 1. Introduction}

Explicit solutions of the Einstein equations can only be found
under restrictive assumptions such as high symmetry. The physical
significance of these solutions depends on the existence of
a class of solutions which is general enough to have a direct
physical application and whose qualitative behaviour resembles
that of the exact solutions. These remarks apply in particular
to the homogeneous and isotropic models used in cosmology. Thus it
is of interest to know which properties of these models are
stable under small perturbations. There is a large literature
concerned with linearized perturbations of such models but this
can at best give a rough indication of the behaviour of small
but finite perturbations. (For a general discussion of the
relations between different concepts of stability see [9].)
The basic question studied in this paper is the following.
Given a homogeneous cosmological model, assumed for simplicity
to have compact spatial sections, what is the behaviour of
solutions which develop from initial data which are, in an
appropriate sense, close to being homogeneous? Do they remain
almost homogeneous? Do they perhaps in some sense become more
homogeneous in the course of time?

At present no rigorous results exist on the question just
raised concerning non-vacuum solutions of the Einstein equations
with almost homogeneous initial data. It appears that the only
results on the nonlinear stability of any solution of the
Einstein equations are those of Friedrich [7] for de Sitter space
and Christodoulou and Klainerman [4] for Minkowski space. In this
paper we study the analogous question for Newtonian cosmological
models. There are at least two motivations for doing this. Firstly,
Newtonian cosmological models are of interest for applications since
they are used to model structure formation in the early universe [15].
Secondly, they may provide useful insights for the relativistic case.

As matter model we take either an isentropic perfect fluid with
polytropic equation of state or dust. The models considered expand
for ever and have positive cosmological constant $\Lambda$. (Some
special solutions with $\Lambda=0$ are also considered for comparison.)
The main results concern those solutions with isotropic
expansion. (The meaning of this terminology will be explained below.)
In the case of a positive cosmological constant it is shown that
the solutions corresponding to nearly homogeneous initial data
exist for all positive times and that the difference between the
inhomogeneous and homogeneous solutions tends to zero in a strong
sense as $t\to\infty$. On the other hand it is shown by an example
using dust that the quantity $(\rho-\bar\rho)/\bar\rho$ often does
not tend to zero. Here $\rho$ denotes the density and $\bar\rho$
the mean density. Another example using dust shows that when the
cosmological constant is zero there are solutions which develop
singularities in finite time and which evolve from data which
are arbitrarily close to homogeneity.

To put these results in context it is useful to compare with
the case of a non-relativistic fluid without gravitation. In
that case the spatially homogeneous solutions are time independent.
It has been shown by Sideris[17] that if the fluid is polytropic
rather general small perturbations of homogeneous initial
data lead to singularities in finite time. Intuitively, one of
the mechanisms leading to singularities in a fluid with
non-vanishing pressure is the formation of shock waves. In the case
of dust shell-crossing singularities will often occur. The local
effects of gravitation cannot be expected to improve this situation
and may even aggravate it. There is, however, a global effect which
can be helpful. This is that the background solution may be expanding
and that this expansion may \lq pull apart\rq\ potential shock
waves and shell-crossing singularities, thus preventing them from
forming. Our results show that when $\Lambda=0$ the expansion is not
powerful enough to achieve this but when $\Lambda>0$ it is, at least
for small initial perturbations. Another comparison which is of some
interest is that with Newtonian cosmological models where the matter
is collisionless and described by the Vlasov equation. This kind
of matter does not have the same kind of tendency to form singularities
as a fluid and it is possible to prove global existence without any
restriction on the size of the perturbations[16]. On the other hand less
is known about the asymptotic properties of the solutions in that case
and it is possible that the existing estimates for the asymptotic
behaviour are not optimal.

The method of proof of the existence theorem is inspired by the
following observation, which is of more general interest.
Consider the symmetric hyperbolic system
$$A^0(u)\d_tu+A^i(u)\d_i u=0\eqno(1.1)$$
for a vector-valued function $u$ in $n$ space dimensions.
One solution of this equation is given by $u=0$. It is a well
known fact (see e.g. [10, 13]) that even arbitrarily small initial
data for (1.1) often leads to a solution which develops a
singularity in finite time. In this sense the solution $u=0$
is unstable. If, however, this equation is replaced by
$$A^0(u)\d_t u+A^i(u)\d_i u+ku=0,\eqno(1.2)$$
where $k$ is a positive constant then the situation changes
dramatically. (This is shown explicitly for the simple scalar
equation $u_t+uu_x+u=0$ in [10].)
The solutions corresponding to any initial
data for (1.2) which are small enough exist for all positive
time and, in fact, converge to zero exponentially as $t\to\infty$.
This can be proved using the fact that energy estimates for
(1.2) show that as long as the solution is small its norm in
the Sobolev space $H^s$, $s\ge 4$, is decreasing and a bootstrap
argument. Below a proof of this kind in a situation which is
slightly more complicated (due to the presence of the Newtonian
potential term) will be presented in detail. In the cosmological
problem the role of the constant $k$ is played by $\dot a/a$ where $a$
is the scale factor of a homogeneous model. For the argument to
work this quantity must be bounded away from zero (or at least
tend to zero sufficiently slowly as $t\to\infty$) and it is
the positive cosmological constant which ensures this. There
is also a term representing the perturbed gravitational field
which must be controlled.

The starting point for the study of Newtonian cosmology in
this paper is the Newton-Cartan theory, which is a geometrical
version and at the same time a slight generalization of
ordinary Newtonian gravitation theory. It is shown that we
can split off from the full equations a system of ordinary
differential equations which describe the evolution of the
mean density, a flat metric and a connection on a torus.
These are identical to the equations which describe the evolution of
spatially homogeneous Newtonian cosmological models.
This means that it is possible
to associate uniquely to each solution of the Newton-Cartan
theory with compact spatial sections a \lq background
solution\rq\ which is spatially homogeneous. It is then useful
for the analysis of the dynamics to regard the given solution
as a perturbation of the corresponding background. In the
literature on Newtonian cosmology it is usual to split
general solutions into the sum of a background and a
perturbation. The analysis in this paper clarifies the
significance of a splitting of this kind and shows that it can be
invariantly defined. This contrasts with the situation in
general relativity where the problem of associating a homogeneous
cosmological model with an inhomogeneous one so as to give a
mathematical formulation of the idea that the universe is on
average homogeneous is a difficult one[6].
The \lq solutions with isotropic expansion\rq\
referred to above are those where the evolution of the flat metric
on the torus is just a rescaling by a spatially constant conformal
factor. If the expansion is not isotropic then certain parts of
the geometry can be given arbitrarily and are not fixed by the
field equations. Thus, as we will see shortly,
the field equations of the Newton-Cartan
theory, unlike the Einstein equations, are not strong enough to
determine a solution uniquely in terms of initial data.

The results of this paper on the global dynamics of Newtonian
cosmological models are linked to two questions of interest in
cosmology. The first is the cosmic no-hair conjecture. This
says that in a cosmological model with positive cosmological
constant perturbations should be strongly damped. This conjecture
refers originally to an inflationary period in general relativity.
It is interesting to consider whether the analogous statement
is true in Newtonian theory. In particular this may give insight
into the mathematical mechanism responsible for the damping.
Because of the quantities which can be given freely in the case
where the expansion is not isotropic, it is clear that not all
perturbations can be damped in the Newtonian case. This fact has
been remarked upon by Barrow and G\"otz [1]. (These authors did
not use the Newton-Cartan theory.) Our results show that in the
isotropic case small perturbations are strongly damped and at the
same time set limits on how strong the damping can be. The second
question is that of structure formation in cosmology. This has
been studied in the context of Newtonian cosmology with a positive
cosmological constant and dust as a matter model by Bildhauer,
Buchert and Kasai[2]. They use the formation of singularities in
a dust solution as a signal for structure formation in a more
realistic model. Their results are consistent with those of the
present paper and indicate that singularities are likely to occur
in finite time for large initial perturbations.

The paper is organized as follows. Section 2 describes the
Newton-Cartan theory and sets up the basic equations needed in the
sequel. In section 3 the results on global existence and
asymptotic behaviour are proved. The examples which show the
sharpness of these results are presented in section 4.

\vskip .5cm
\noindent
{\bf 2. Newtonian cosmology}

In this section we present some general information about spatially
compact cosmological models in the Newtonian theory of gravity. More
precisely, since Newtonian gravitation as it is normally presented
is only applicable to isolated systems, we use a slight generalization
of the standard Newtonian formalism, the Newton-Cartan theory. The
original treatment of Newtonian cosmology, due to Heckmann and
Sch\"ucking [8], is more elementary but does not bring out the
underlying geometrical structures. The
definition which will be used here is that which has been discussed
by Ehlers[5] in the context of the Newtonian limit of general
relativity. The basic objects of the theory are a covariant symmetric
tensor $g_{\alpha\beta}$ (the time metric), a contravariant tensor
$h^{\alpha\beta}$ (the space metric), a torsion-free connection
$\Gamma^\alpha_{\beta\gamma}$ and a symmetric tensor $T^{\alpha\beta}$
(the matter tensor) defined on a four-dimensional manifold (spacetime).
They are supposed to satisfy the following conditions:

\noindent
1) At each point of spacetime there exists at least one vector
$V^\alpha$ with $g_{\alpha\beta}V^\alpha V^\beta>0$. Vectors with
this property are called timelike.

\noindent
2) If $V^\alpha$ is timelike then the restriction of $h^{\alpha\beta}$
to the space of covectors $\omega_\alpha$ with
$\omega_\alpha V^\alpha=0$ is positive definite.

\noindent
3) $g_{\alpha\beta}h^{\beta\gamma}=0$.

\noindent
4) $g_{\alpha\beta}$ and $h^{\alpha\beta}$ are covariantly constant
with respect to $\Gamma^\alpha_{\beta\gamma}$.

\noindent
5) The curvature tensor $R^\alpha_{\beta\gamma\delta}$ has the
symmetry property $h^{\sigma\gamma}R^\alpha_{\beta\sigma\delta}=
h^{\sigma\alpha}R^\gamma_{\delta\sigma\beta}$.

\noindent
6) $R_{\alpha\beta}=8\pi
(g_{\alpha\sigma}g_{\beta\tau}T^{\sigma\tau}
-{1\over 2} g_{\sigma\tau}T^{\sigma\tau}g_{\alpha\beta})
-\Lambda g_{\alpha\beta}$ where $R_{\alpha\beta}$ is the Ricci
tensor of $\Gamma^\alpha_{\beta\gamma}$ and $\Lambda$ is the
cosmological constant.

\noindent
7) $\nabla_{\alpha}T^{\alpha\beta}=0$.

\noindent
It is a known consequence of these conditions that the 3-dimensional
distribution defined as the kernel of $g_{\alpha\beta}$ is integrable.
The integral manifolds are the surfaces of constant absolute time $t$. A
spatially compact cosmological model in this theory can be defined as
one where the hypersurfaces of constant time are compact. The tensor
$h^{\alpha\beta}$ defines a Riemannian metric on each time slice and
the field equation (condition 6) above) implies that this metric is
flat. This means in particular that if a time slice is compact it
must be isometric to a quotient of a flat torus by a finite group
of isometries[12].

By analogy with general relativity, a Newtonian spacetime $M$
will be called time-orientable if there exists a global smooth
timelike vector field $V^\alpha$. It can be assumed without loss of
generality that $g_{\alpha\beta}V^\alpha V^\beta=1$. Let $S$
be one of the time slices and define a smooth mapping $f$ from an
open subset $U$ of $S\times\R$ to $M$ by the condition that
$f(x,s)$ is the point of $M$ obtained by starting at $x$ and
following an integral curve of $V^\alpha$ for a parameter
distance $s$. The set $U$ is supposed to be the maximal one
for which such a mapping is defined i.e. each integral curve
is followed to the end. The normalization condition ensures
that the hypersurfaces $t$=const. coincide with the images of
the hypersurfaces $s$=const. under $f$. This in turn implies that
$U$ is of the form $S\times I$ for some open interval $I$,
since otherwise the time slices would fail to be compact.
The mapping $f$ is a local diffeomorphism. In particular $f(U)$
is open. This construction can be started from any slice and
so if $f$ were not onto it would be possible to write $M$ as
a disjoint union of non-empty open sets. If spacetime is assumed
to be connected this is impossible and so $f$ is surjective.
To get the injectivity of $f$ it seems to be necessary to make
another causality assumption. Call a subset of a Newtonian
spacetime {\it achronal} if there exists no timelike curve
starting and ending there. The following lemma can now be stated.

\noindent
{\bf Lemma 2.1}\next
{\it  Let $M$ be a Newtonian spacetime with the
following properties:

\noindent
(i) it is time orientable

\noindent
(ii) each time slice is compact and achronal

\noindent
Then $M$ is diffeomorphic to $S\times I$, where $S$ is a compact
manifold admitting a flat metric and $I$ is an open interval.
}

\vskip 10pt

\noindent
As already remarked, such a manifold $S$ is covered by a torus.
Passing to this covering does not change the dynamics and so
when investigating dynamical questions it may be assumed without
loss of generality that $S$ is a torus. This will be done in
the following.

Represent $S$ in the form $\R^3/\sim$, where the projection from
$\R^3$ to $S$ is smooth. The tensor $h_{ab}$ on $M$ defines
a tensor on $\R^3\times I$ which will be denoted in the same
way. Let $h_{ab}(t)$ be its restriction to $\R^3\times\{t\}$.
This is a complete flat metric on $\R^3$ and so must be isometric
to the standard metric on $\R^3$[12]. This implies that given any
orthonormal frame at the origin there exist coordinates on
$\R^3\times\{t\}$ such that the metric takes the form $\delta_{ab}$
and the given frame takes the form $\d/\d x^a$. Introduce coordinates
of this type on each slice. Since these can be constructed by
following geodesics the $x^a$ are smooth functions on $\R^3\times I$.
In these coordinates the components $h_{ab}=\delta_{ab}$.
The identification which leads from $\R^3$ to $S$ is of the form
that $x^a$ is identified with $x^a+n^ie_i^a(t)$ for each triple of
integers $n^i$. What is not a priori clear is that $e^a(t)$ is smooth
(or even continuous). A covering transformation of $\R^3\times I$ is
smooth and must be of the form $x^a\mapsto x^a+n^ie_i^a(t)$ for
each fixed $t$. This is only possible if $e_i^a(t)$ is smooth. Doing
a time dependent linear tranformation of the coordinates leads to a
coordinate system where the identifications are time independent
and the components $h_{ab}$ only depend on $t$.

The picture is now that the spacetime is the product of a torus
with an interval and there are periodic coordinates such that
the time and space metrics take the form
$$\left.\eqalign{ &g_{00}=1,\ g_{0a}=0,\ g_{ab}=0,\cr
&h^{0\alpha}=0,\  h^{ab}=h^{ab}(t).}\right\}\eqno(2.1)$$
Certain restrictions on $\Gamma^\alpha_{\beta\gamma}$ follow from
(2.1) and assumption 4) above. They are
$$\left.\eqalign{&\Gamma^0_{\alpha\beta}=0,\ \Gamma^a_{bc}=0,\cr
&\partial_t h^{ab}+\Gamma^a_{s0}h^{sb}+\Gamma^b_{s0}h^{sa}=0}\right
\}\eqno(2.2)$$
Names will now be assigned to the parts of the connection which
do not vanish as a consequence of (2.2).
$$\eqalignno{&E^a=\Gamma^a_{00},&(2.3)\cr
&\Theta_{ab}=\f12(h_{ac}\Gamma^c_{b0}+h_{bc}\Gamma^c_{a0}),&(2.4)\cr
&\Omega_{ab}=\f12(h_{ac}\Gamma^c_{b0}-h_{bc}\Gamma^c_{a0}).&(2.5)}$$
Combining (2.2) and (2.4) gives
$$\partial_t h_{ab}=2\Theta_{ab}
.\eqno(2.6)$$
This means in particular that $\Theta_{ab}$ only depends on $t$ and
not on the spatial coordinates. Let $\rho=T^{00}$. Condition 6) gives
the equations
$$\eqalignno{
&\nabla_a E^a=h^{ab}\d_t\Theta_{ab}-\Theta^{ab}\Theta_{ab}-\Omega^{ab}
\Omega_{ab}+4\pi \rho-\Lambda,&(2.7) \cr
&h^{bc}\nabla_c\Omega_{ab}=0,&(2.8)}$$
where indices have been raised with $h^{ab}$ while condition 5)
implies the equations
$$\eqalignno{
&\nabla_{[c}\Omega_{ab]}=0, &(2.9) \cr
&\d_t \Omega_{ab}=\nabla_{[b} E_{a]}. &(2.10)}$$
Combining (2.8) and (2.9) shows that $\Omega_{ab}$ is also
independent of the spatial coordinates. Integrating (2.10) over
the torus and applying Stokes' theorem gives
$$\d_t\Omega_{ab}=0.\eqno(2.11)$$
Substituting this back into (2.10) shows that the one-form $E_a$ is
closed. Thus
$$E_a=\eta_a-\nabla_a U,\eqno(2.12)$$
where $\eta_a$ depends only on time and the function $U$, which is
determined only up to addition of a function of $t$ by (2.12), is
supposed to satisfy $\int U=0$. Consider now the effect of doing
a coordinate transformation of the form ${x'}^a=x^a+y^a$, $t'=t$
where the $y^a$ depend only on $t$. This preserves the conditions
(2.1) and changes $E_a$ by $h_{ab}\ddot x^b$. Thus it can be used
to eliminate $\eta^a$ without disturbing the assumptions made up
to now. It follows that it can be assumed without loss of generality
that $\eta^a=0$. Before (2.7) is analysed $\Theta_{ab}$
will be split into trace and tracefree parts:
$$\Theta_{ab}=\Sigma_{ab}+\f13\theta h_{ab},\eqno(2.13)$$
where $h^{ab}\Sigma_{ab}=0$ and $\theta=h^{ab}\Theta_{ab}$. Then (2.7)
becomes
$$\Delta_h U=-[\dot\theta+\f13\theta^2+\Sigma_{ab}
\Sigma^{ab}-\Omega_{ab}\Omega^{ab}+4\pi\rho-\Lambda]\eqno(2.14)$$
where $\Delta_h U=h^{ab}\nabla_a\nabla_b U$.
Let $\bar\rho(t)$ denote the average density at time $t$, i.e.
$\bar\rho(t)=\int \rho(t,x) dx/\int dx$. Then
averaging (2.14) over the torus and using the fact that $\theta$,
$\Sigma_{ab}$ and $\Omega_{ab}$ are spatially constant gives
$$\dot\theta+\f13\theta^2+\Sigma_{ab}\Sigma^{ab}-\Omega_{ab}\Omega^{ab}
+4\pi\bar\rho-\Lambda=0.\eqno(2.15)$$
Putting this back in (2.14) shows that
$$\Delta_h U=-4\pi (\rho-\bar\rho).\eqno(2.16)$$
Thus the Poisson equation has been recovered in a familiar form.

It will now be discussed how the solutions of the equations can
usefully be parametrized. Taking the zeroth component of the
equation $\nabla_\alpha T^{\alpha\beta}=0$ of condition 7) and
averaging over the torus gives
$$\d_t\bar\rho+\theta\bar\rho=0.\eqno(2.17)$$
To specify a solution we need to give the following objects: the
matter variables (which in particular allow $\rho$ to be calculated),
$\theta$ and $\Omega_{ab}$ on the initial hypersurface $t=0$ and
$\Sigma_{ab}$ everywhere. Then, using (2.6), (2.11),
(2.15) and (2.17) the quantities $h_{ab}$, $\bar\rho$, $\theta$
and $\Omega_{ab}$ are determined everywhere. All Christoffel symbols
can be calculated from $h_{ab}$, $\theta$, $\Sigma_{ab}$,
$\Omega_{ab}$ and $\nabla_a U$. The only one of these
which is not already known is $\nabla_a U$. For reasonable matter
fields the system of matter fields coupled to (2.16) should have a
well posed initial value problem. In the case of a perfect fluid the
equations look as follows:
$$\eqalignno{&\d_t\rho+\nabla_a(\rho u^a)+\theta\rho=0&(2.18)   \cr
&\d_t u^a+u^b \nabla_b u^a+\rho^{-1}h^{ab}\nabla_b p+\f23
\theta u^a-\nabla^a U
+2(\Sigma^a_b+\Omega^a_b)u^b=0&(2.19)}$$
assuming that the density $\rho$ does not vanish anywhere. These
equations are obtained by choosing the matter tensor
$$T^{\alpha\beta}=\rho u^\alpha u^\beta+ ph^{\alpha\beta}\eqno(2.20)$$
with $u^0=1$ and computing the condition 7) explicitly. That this
procedure does in fact work for a fluid will be shown in the next
section (Theorem 3.1).

The simplest solutions of these equations are the Friedmann-like
solutions which are obtained by taking $\rho=\bar\rho$, $u^a=0$,
$E_a=0$, $\Sigma_{ab}=0$ and $\Omega_{ab}=0$. In this case (2.6) reduces
to $\d_t h_{ab}=\f23\theta h_{ab}$, which implies that $h_{ab}
=a^2(t)h_{ab}(0)$ for some function $a(t)$. Suppose without loss
of generality that $a(0)=1$. This function satisfies
the equation $\dot a=\f13\theta a$. Substituting this into (2.17)
gives
$$\rho a^3=\rho(0).\eqno(2.21)$$
Combining (2.15), (2.17) and (2.21) results in the equation
$$\ddot a=-\f{4\pi}3\rho(0)a^{-2}+\f13\Lambda a.\eqno(2.22)$$
For solutions of (2.22) the quantity $E=\f12\dot a^2-\f{4\pi}3\rho(0)a^{-1}
-\f16\Lambda a^2$ is constant and so the qualitative behaviour of
its solution can easily be analysed using an effective potential.
(The equation (2.22) also describes homogeneous and isotropic dust
solutions in general relativity and in that case $E=-k/2$ where
$k$ is the curvature of the space sections.)
This analysis shows that in the case $\Lambda>0$, which is the one
of greatest interest in the following, $a$ tends to infinity as
$t\to\infty$ precisely in the cases:

\noindent
(i) $E>-\f12(4\pi)^{2/3}\Lambda^{1/3}\rho(0)^{2/3}$

\noindent
(ii) $E\le -\f12(4\pi)^{2/3}\Lambda^{1/3}\rho(0)^{2/3}$ and
$a(0)>(4\pi\rho(0)/\Lambda)^{1/3}$

\vskip 10pt\noindent
The constancy of $E$ implies that for these solutions $\dot a/a\to
\sqrt{\Lambda/3}$ as $t\to\infty$ so that asymptotically the
expansion becomes exponential.

\vskip .5cm
\noindent
{\bf 3. The main theorems}

Now general solutions of the equations of Newtonian cosmology
of the previous section with a perfect fluid as a matter model
will be considered. The first question to be asked is whether
the system of equations consisting of (2.16), (2.18) and (2.19)
has a well posed initial value problem. There are various ways in
which (2.18) and (2.19) can be written as a symmetric hyperbolic
system for fixed $U$ if an equation of state $p=f(\rho)$ with
$f'>0$ is given. In the following only polytropic equations of
state will be considered i.e. $p=K\rho^{n+1\over n}$, where $K$
and $n$ are positive real numbers. It will be convenient to
use the variable $w=\sqrt{2n(2n+2)K}\rho^{1/2n}$ introduced by
Makino [14]. Equations (2.18) and (2.19) can then be written as follows.
$$\eqalignno{
&\d_t w+u^a\d_a w+\f1{2n}w\d_a u^a+\f1{2n}\theta w=0&(3.1)     \cr
&\d_t u^a+u^b\d_b u^a+\f1{2n}wh^{ab}\d_b w-h^{ab}
\nabla_b U+\f23\theta u^a+2(\Sigma^a{}_b+\Omega^a{}_b)u^b=0
&(3.2)}$$
The term in this system involving $U$ is $-\nabla^a U$ and this can be
rewritten formally as $4\pi\nabla^a\Delta_h^{-1}(\rho-\bar\rho)$.
This should be thought of as a lower order term, since it is formally
of order -1. Accepting this for the moment, we see that contracting
(3.2) with $h_{ac}$ and taking the result together with (3.1) gives
a system which is exactly like a symmetric hyperbolic system except
that it contains one lower order term which is non-local. It is then
natural to prove local in time existence for (2.16), (2.18) and (2.19)
by following the steps in the standard local existence proof for
symmetric hyperbolic systems as given in [13] for instance and checking
that the non-local term does not cause any problems. This has been done
in [3] for the case $\Sigma_{ab}=0$, $\Omega_{ab}=0$.
The general case is not significantly different. There results the
following theorem:

\noindent
{\bf Theorem 3.1} \next
{\it Let $\Sigma_{ab}$ be a given $C^\infty$
function of $t$. Let $\Omega^0_{ab}$ be an antisymmetric matrix,
$h^0_{ab}$ a symmetric matrix, $\theta_0$ a real number
and $\rho_0$, $u^a_0$ functions in $H^s(T^3)$ for some $s\ge3$.
Suppose that $\rho$ is positive.
Then there exists a $T>0$ and a unique solution of equations (2.6)
(2.11), (2.15), (2.16), (2.18) and (2.19) on $[0,T)$ such that
\next
(i) $\Omega_{ab}$, $h_{ab}$ and $\theta$ are $C^\infty$ functions
of $t$ which take on the values $\Omega^0_{ab}$, $h^0_{ab}$ and
$\theta_0$ respectively for t=0
\next
(ii) $\rho$ and $u^a$ are $C^0$ with values in $H^s(T^3)$ and
$C^1$ with values in $H^{s-1}(T^3)$ and take on the values
$\rho_0$ and $u^a_0$ respectively for t=0
\next
Moreover for this solution:
\next
(iii) $U$ is $C^0$ with values in $H^{s+1}(T^3)$ and $C^1$
with values in $H^s(T^3)$
\next
(iv) if $\rho$ is uniformly bounded away from zero and the $C^1$
norms of $\rho$ and $u^a$ are bounded on the interval $[0,T)$
then the solution can be extended to an interval $[0,T_1)$ with
$T_1>T$
}
\vskip 10pt
Next the global behaviour of solutions with almost homogeneous
initial data will be discussed. At this stage it will be assumed
that $h_{ab}(0)=\delta_{ab}$, $\Sigma_{ab}=0$ and $\Omega_{ab}=0$.
Some remarks will be made later on what happens when
these quantities are allowed to be non-zero. Under these
restrictions the spatial metric can be written in the form
$h_{ab}(t)=a^2(t)\delta_{ab}$. Then $\theta=3\dot a/a$. The equations
(3.1) and (3.2) simplify and it is possible to obtain a symmetric
hyperbolic system from them in a slightly different way from what
was done previously. If $v^a=au^a$ is taken as a new variable then
the system is symmetric hyperbolic without contracting (3.2) with
the metric. This has the advantage that the coefficient matrix of
the time derivatives in the resulting system is the identity and
this simplifies the derivation of energy estimates. Writing $V$
for the pair $(w-\bar w, v^a)$, the system is now of the form
$$\d_t V+A^a\d_a V+BV+PV=0.\eqno(3.3)$$
Here $B={\rm diag}\{3\dot a/2na, \dot a/a, \dot a/a, \dot a/a\}$
and $PV=(0,aP_1V)$, where $P_1V$ is the gradient of the potential
$U$ of the density corresponding to $w$. All the matrices $A^i$
have a similar form. For example
$$A^1=\left(\matrix{ v^1/a&w/2na&0&0\cr
                     w/2na&v^1/a&0&0\cr
                     0&0&v^1/a&0\cr
                     0&0&0&v^1/a}\right)$$

In deriving estimates for these equations we will repeatedly
use the following Moser inequalities (see [11] for a derivation).
$$\eqalign{
\|D^\alpha(fg)\|_2&\le C(\|f\|_\infty\|D^sg\|_2
+\|D^sf\|_2\|g\|_\infty)             \cr
\|D^\alpha(fg)-fD^\alpha g\|_2&\le C(\|Df\|_\infty\|D^{s-1}g\|_2
+\|D^s f\|_2\|g\|_\infty)            \cr
\|D^\alpha(F(f))\|_2&\le C|\d F/\d f|_{C^{s-1}}\|f\|_\infty^{s-1}
\|Df\|_\infty\|D^s f\|_2}$$
Here the subscripts 2 and $\infty$ refer to the $L^2$ and $L^\infty$
norms respectively, $\alpha$ is a multi-index and $s=|\alpha|$.

The basic energy estimate for (3.3) goes as follows:
$$\eqalign{
{d\over dt}\|V\|_2^2&=2\int\langle V,\d_t V\rangle dx    \cr
                  &=-2\int\langle V, A^i\d_iV+BV+PV\rangle dx \cr
                  &\le (\|\d_iA^i\|_\infty-2\min(\f3{2n},1)\f{\dot a}
a)\|V\|_2^2+2\|V\|_2\|PV\|_2}\eqno(3.4)$$
Now apply a spatial derivative $D^\alpha$ ($\alpha$ a multi-index)
to (3.3) to get
$$\d_t(D^\alpha V)+A^i\d_i(D^\alpha V)+BD^\alpha V+D^\alpha PV=
-[D^\alpha(A^i\d_i V)-A^iD^\alpha(\d_i V)]\eqno(3.5)$$
Applying the Moser estimates to the term on the right hand side
gives
$$\|D^\alpha(A^i(V)\d_iV)-A^i(V)D^\alpha(\d_i V)\|_2\le
C|\d A/\d V|_{C^{s'}}(1+\|V\|_\infty^{s-1})\|DV\|_\infty\|D^s V\|_2
\eqno(3.6)$$
where $s=|\alpha|$ and $s'=\max (1,s-1)$. Thus
$$\eqalign{&{d\over dt}
\|D^s V\|_2^2\le [C|\d A/\d V|_{C^{s'}}(1+\|V\|_\infty^{s-1})
\|DV\|_\infty-2\min(\f3{2n},1)\f{\dot a}a]\|D^s V\|_2^2     \cr
&\qquad+2\|D^sV\|_2\|D^s(PV)\|_2.}\eqno(3.7)$$
It remains to estimate $PV$. this will be done not in terms of $w$ but
in terms of $\rho-\bar\rho$. Standard elliptic theory gives the
estimate
$$\|D^s(PV)\|_2\le Ca\|\rho-\bar\rho\|_{H^{s-1}}\eqno(3.8)$$
This inequality is only helpful if we have an estimate for
$\rho-\bar\rho$ independent of (3.7). To obtain this, write the
continuity equation in terms of $a^k(\rho-\bar\rho)$, where $k$ is
some number $\le 3$ to be fixed later. The result is
$$\d_t(a^k(\rho-\bar\rho))+a^ku^i
\nabla_i(\rho-\bar\rho)-{(k-3)\dot a\over a}a^k(\rho-\bar\rho)=a^k\rho
\nabla_i u^i\eqno(3.9)$$
Multiplying this by $a^k(\rho-\bar\rho)$ gives an energy type estimate
for $\rho-\bar\rho$. The equation (3.9) can be differentiated in space
to get higher energy estimates. The final result is
$$\eqalign{&{d\over dt}
\|a^kD^{s-1}(\rho-\bar\rho)\|_2^2\le
(C\|Du\|_\infty+\f{(k-3)\dot a}a)\|a^kD^{s-1}(\rho-\bar\rho)\|_2^2  \cr
&\qquad+C[\|D(a^k(\rho-\bar\rho))\|_\infty
+\|a^k(\rho-\bar\rho)\|_\infty+\|a^k\bar\rho\|_\infty]
\|a^kD^{s-1}(\rho-\bar\rho)\|_2\|D^s u\|_2.}
\eqno(3.10)$$
The inequalities (3.7) and (3.10) are almost what is needed for
the global existence theorem but still need to be altered slightly.
First note that the expression
$C|\d A/\d V|_{C^{s'}}(1+\|V\|_\infty^{s-1})$
can be thought of as a constant depending continuously on
$\|V\|_\infty$. Next, the dynamics of the scale factor $a$
implies that there exists a constant $H_0>0$, only depending on the
initial data, such that $\dot a/a\ge H_0$ as long as a solution
exists. It is also obviously possible to replace $L^2$ norms of
derivatives of order $s$ of $V$ and $PV$ by $H^s$ norms of these
functions themselves. The $H^s$ norm of $PV$ can be eliminated in
favour of $\rho-\bar\rho$ using (3.8). Finally the whole inequality
can be divided by $\|V\|_{H^s}$. Hence
$${d\over dt}\|V\|_{H^s}\le
[C(\|V\|_\infty)\|DV\|_\infty-\min(\f3{2n},1)H_0]\|V\|_{H^s}+
Ca^{1-k}\|a^k(\rho-\bar\rho)\|_{H^{s-1}}\eqno(3.11)$$
In (3.10) the quantity $H_0$ can be introduced in the same way.
The inequality
can be divided by $\|a^k(\rho-\bar\rho)\|_{H^{s-1}}$, giving
$$\eqalign{
{d\over dt}\|a^k(\rho-\bar\rho)\|_{H^{s-1}}&\le [C\|DV\|_\infty+
(k-3)H_0]\|a^k(\rho-\bar\rho)\|_{H^{s-1}}       \cr
&\qquad+\|V\|_{H^s}[\|D(a^k(\bar\rho-\rho))\|_\infty
+\|a^k(\rho-\bar\rho)\|_\infty+\|a^k\bar\rho\|_\infty].}
\eqno(3.12)$$
The inequalities (3.11) and (3.12) are the main tools needed to
prove the main theorem.

\noindent
{\bf Theorem 3.2} \next
{\it Let initial data be given for (3.3) as in
Theorem 3.1 with $s\ge 4$. If $a_0$, $\theta_0$, $\bar\rho_0$
and $\Lambda>0$ are fixed then there exists a positive number
$\epsilon$ such that if
$$\|\rho_0-\bar\rho_0\|_{H^s}+\|v_0\|_{H^s}<\epsilon\eqno(3.13)$$
the corresponding solution (whose local existence is guaranteed by
Theorem 3.1) exists globally in time. Moreover the quantity
$\|v\|_{H^s}$ converges exponentially to zero and
$(\rho-\bar\rho)/\bar\rho$ tends to a limit in ${H^{s-2}}$
as $t\to\infty$.
}

\noindent
{\bf Proof} \next
Choose some $k$ lying strictly between 1 and 3 and
define $X=\|a^k(\rho-\bar\rho)\|_{H^{s-1}}+\|V\|_{H^s}$. Let
$h=\min((3-k)H_0,3/2n,1)$. Then it follows from (3.11) and (3.12)
that
$${dX\over dt}\le [C(X)X-h]X+Ca^lX.\eqno(3.14)$$
Here $C(X)$ can be chosen to depend continuously on $X$ (and hence
to be bounded on bounded subsets) and $l=\max(1-k, k-4)$.
The assumption $s\ge 4$ is necessary so that
the expression $\|D(a^k(\rho-\bar\rho))\|_\infty$ occurring in
(3.12) can be estimated by $\|a^k(\rho-\bar\rho)\|_{H^{s-1}}$.
The essential idea of the proof is now as follows. If $X$ is
sufficiently small then the first term on the right hand side
of (3.14) will be negative and so will tend to cause $X$ to
become smaller. However more information is needed in order to
conclude that $X$ actually gets smaller since this term must
be negative enough to outweigh the effect of the second term.
This will be the case if $t$ is large enough (and hence $a^l$
is small enough). Thus it is appropriate to do a bootstrap
argument on a sufficiently long time interval.

Let $\delta_1$ be a positive number which is small enough so that
$X<\delta_1$ implies that $C(X)X<h/3$. If for a given solution of
(3.3) on an interval $[0,T)$ the quantity $X$ is bounded then the
solution can be extended to a larger time interval. Thus the
inequality (3.14) shows that given any $T_1>0$ there is some
$\delta_2>0$ such that if $X$ is less than $\delta_2$ for $t=0$ the
solution exists up to time $T_1$ and $X<\delta_1$ on the whole
interval $[0,T_1]$. Now choose $T_1$ so that $Ca^l(t)<h/3$ for all
$t\ge T_1$. (This is possible since $a\to\infty$ as $t\to\infty$.)
Choose $\epsilon>0$ so that the inequality (3.13) implies that the
initial value of $X$ is smaller than $\delta_2$. Now consider an
initial datum for which (3.13) is satisfied. Let $T^*$ be the
largest time (possibly infinite) for which the corresponding solution
exists on $[0,T^*)$ and $X\le \delta_1$ for $t<T^*$. Clearly
$T^*>T_1$. On the interval $[T_1,T^*)$ (3.14) implies that
$dX/dt\le -\f13 h X$. It follows in particular that $X$ is strictly
decreasing there. Hence $X(t)<X(T_1)$ on $[T_1,T^*)$. If $T^*$
were finite we could conclude firstly that the solution extends
to a longer time interval (since $X$ is bounded) and secondly that
the inequality $X(t)\le \delta_1$ holds on a longer time interval.
The second point follows from the fact that $X(t)<X(T_1)<\delta_1$
on $[T_1,T^*)$ and continuity. This contradicts the definition of
$T^*$. The conclusion is that $T^*=\infty$. So the solution
exists globally and $dX/dt\le -\f13 hX$ for ever. A simple
comparison argument then shows that $X(t)\le X(T_1)\exp (-\f13 ht)$
for all $t\ge T_1$.
Going back to the definitions shows that $\|v\|_{H^s}=a\|u\|_{H^s}$
is $O(exp(-\f13 ht))$ as $t\to\infty$. Since $a$ itself tends to
infinity it follows that $\|v\|_{H^s}$ and $\|u\|_{H^s}$ converge to
zero exponentially as $t\to\infty$. To get the remaining statement
of the theorem, consider the inequality (3.12) in the case $k=3$.
Writing $Y=\|(\rho-\bar\rho)/\bar\rho)\|_{H^{s-1}}$ this leads
to an estimate of the form
$${dY\over dt}\le CX(Y+1)\eqno(3.15)$$
Applying Gronwall's inequality and using the fact that
$\int_0^\infty X(t)dt<\infty$ shows that $Y$ is bounded. Using
equation (3.9) in the case $k=3$ gives the inequality
$$\eqalign{
\|{(\rho-\bar\rho)\over\bar\rho} (t_2)-{(\rho-\bar\rho)\over\bar\rho}
(t_1)\|_{H^{s-2}}&\le C\int_{t_1}^{t_2}\|V\|_{H^s} dt      \cr
&\le Ce^{-\f13 h(t_2-t_1)}}$$
for any two times $t_1$ and $t_2$ with $t_1<t_2$.
This implies that there exists a function $f$ such that $(\rho-
\bar\rho)/\bar\rho\to f$ in $H^{s-2}$ as
$t\to\infty$.
\BeweisEnde

Theorems 3.1 and 3.2 concern fluids with polytropic equations of
state. Very similar results can be proved for dust using the
same methods. The differences will now be indicated briefly.
Firstly, no Makino variable is necessary in the dust case;
the density can be used directly. Secondly, the equations
(2.18) and (2.19) are only coupled via the potential $U$
when the pressure vanishes. Hence instead of (2.18) and (2.19)
together forming a system which can be written in symmetric
hyperbolic form, the equation (2.19) is symmetric hyperbolic
by itself. Thus $V$ can be taken to be just $v^a$ in this case.
Apart from this the argument works just as for a polytropic
fluid.

If $\Sigma_{ab}$ is arbitrary then the analogue of the proof of
Theorem 3.2 breaks down in general. However it is possible to
generalize somewhat. Firstly it may be observed that $\Omega_{ab}$
makes no contribution in the energy estimates. Hence the analogue
of Theorem 3.2 remains true if an arbitrary $\Omega_{ab}$
is allowed. In the case of non-zero $\Sigma_{ab}$ it turns
out that the condition needed to make the argument go through
is that $\Theta_{ab}$ should be positive definite. To ensure
this it suffices to ensure a sufficiently small uniform bound
on $\Sigma^a_b$. In general relativity it turns out that the
shear, which is analogous to $\Sigma_{ab}$, decays exponentially in
homogeneous cosmological models with positive cosmological
constant[18]. Thus the results found here can be regarded as
indirect positive evidence for the validity of the cosmic no-hair
conjecture in general relativity.

\vskip .5cm
\noindent
{\bf 4. Sharpness of the results}

In this section some examples will be presented which demonstrate
the sharpness of the results obtained in the previous section. They
all concern dust solutions and are proved by studying the paths
of dust particles i.e. the integral curves of $u^a$. The solutions
studied are such that $\rho$ and $u^a$ only depend on $t$ and one
spatial coordinate, which will be denoted by $x$. Moreover it is
asumed that the velocity is of the form $u\d/\d x$ for some
function $u$. Under these assumptions the equations reduce to
$$\left.\eqalign{
\rho_t+(\rho u)_x+\theta\rho&=0                \cr
u_t+uu_x+\f23\theta u-a^{-2}U_x&=0             \cr
U_{xx}=-4\pi a^2(\rho-\bar\rho)}\right\}\eqno(4.1)$$
The equation of motion of a dust particle is $dx/dt=u(x(t),t)$.
Differentiating this with respect to $t$ and using (4.1) gives:
$$d^2 x/dt^2=(-\f23\theta u+a^{-2}U_x)(x).\eqno(4.2)$$
Now consider two integral curves $x_1(t)$ and $x_2(t)$.
$$\eqalign{
{d^2 \over dt^2}(x_2-x_1)&=-{2\over3}\theta{d\over dt}(x_2-x_1)
+\int_{x_1(t)}^{x_2(t)} a^{-2}U_{xx}dx                \cr
&=-{2\over3}\theta {d\over dt}(x_2-x_1)-4\pi\int_{x_1(t)}^{x_2(t)}(\rho-
\bar\rho) dx.}\eqno(4.3)$$
Since $\bar\rho a^3$ is independent of time it follows that
$${d\over dt}\int_{x_1(t)}^{x_2(t)}\bar\rho a^3 dx
=\left(\bar\rho a^3\right)|_{t=0}
\left({dx_2\over dt}-{dx_1\over dt}\right).\eqno(4.4)$$
On the other hand (4.1) and the relation $\theta=3\dot a/a$ give
$$d/dt\int_{x_1(t)}^{x_2(t)}\rho a^3 dx=0,\eqno(4.5)$$
which just expresses the conservation of mass. Thus
$$d^2/dt^2(x_2-x_1)=(Aa^{-3}-2\dot a/a)d/dt(x_2-x_1)-Ba^{-3},\eqno(4.6)$$
for some positive constants $A$ and $B$. If $d/dt(x_2-x_1)$ is
initially non-positive then it will remain so. To see this note that
if it is originally strictly negative it will remain so at least for a
short time by continuity. If it is originally zero then (4.6) shows
that it will immediately become negative and the same conclusion
holds. Now suppose that $d/dt(x_2-x_1)$ becomes zero at some later
time $t_1$. Then the second derivative of $x_2-x_1$ at $t_1$ is
non-negative and this contradicts (4.6). Thus $d/dt(x_2-x_1)$ remains
negative as long as the solution exists.

It will now be shown that for certain solutions of (4.1) the quantity
$(\rho-\bar\rho)/\bar\rho$ cannot converge uniformly to zero as
$t\to\infty$. This shows that the fall-off property of the density
perturbation proved in the last section is optimal. If the initial
datum for $\rho$ is not constant it is possible to choose starting
values for $x_1$ and $x_2$ so that
$$\int_{x_1(0)}^{x_2(0)}\rho-\bar\rho dx>0\eqno(4.7)$$
Now choose the initial velocity to be such that $u(x_2)-u(x_1)\le0$.
Then $u(x_2(t))-u(x_1(t))$ remains non-positive and so
$x_2(t)-x_1(t)\le x_2(0)-x_1(0)$.
Let
$$M(t)=(a(t))^3\int_0^{2\pi}\bar\rho(t) dx= 2\pi(a(t))^2\bar\rho(t)
\eqno(4.8)$$
Then $M(t)$ is constant. Now
$$\eqalign{
\int_{x_1(t)}^{x_2(t)}(\rho-\bar\rho) dx/\bar\rho&>
C(\int_{x_1(t)}^{x_2(t)}(\rho-\bar\rho) dx)a^3              \cr
&=C[(\int_{x_1(t)}^{x_2(t)}\rho dx)a^3-\f1{2\pi}(x_2(t)-x_1(t))M(t)] \cr
&>C[(\int_{x_1(t)}^{x_2(t)}\rho dx)a^3-\f1{2\pi}(x_2(0)-x_1(0))M(0)]}
\eqno(4.9)$$
Hence $\int_{x_1(t)}^{x_2(t)}(\rho-\bar\rho) dx/\bar\rho$ is bounded
below by a positive constant and this is enough to show that
$(\rho-\bar\rho)/\bar\rho$ does not converge uniformly to zero.

The argument just given applies for any value of the cosmological
constant. It will now be shown that the analogues of the results
of section 3 fail when the cosmological constant is zero. First
the solution of (2.22) for $\Lambda=0$ and $E=0$ will be computed.
In that case $\dot a=\sqrt{8\pi\rho(0)/3}a^{-1/2}$ and hence, after
a translation in time, the solution is of the form $a=Kt^{2/3}$. In
particular the solution expands for ever. Now choose data once again
for the fluid and for two dust particles so that $dx_2/dt-dx_1/dt\le0$
as long as the solution exists. Then (4.6) implies that
$$d^2/dt^2(x_2-x_1)\le -\f43 t^{-1}d/dt(x_2-x_1)-BK^{-3}t^{-2}
\eqno(4.10)$$
Consider now the corresponding differential equation
$$d^2 y/dt^2+\f43 t^{-1}dy/dt=-BK^{-3}t^{-2}\eqno(4.11)$$
Its general solution is
$$y=-3BK^{-3}\log t-3C_1t^{-1/3}+C_2\eqno(4.12)$$
The qualitative behaviour of $(x_1-x_2)(t)$ can be studied by
comparing it with the solution $y(t)$ of (4.11) with the same
initial data. Equation (4.11) is a first order ODE for $dy/dt$
and so standard comparison arguments for solutions of first order
ODE's give ${d\over dt}(x_2-x_1)\le {dy\over dt}$. This inequality
can then be integrated again to show that $x_2-x_1\le y$.
Now (4.12) implies that $y\to0$ in finite time if the solution of
(4.1) exists that long. The trajectories of two dust particles in a
regular solution of (4.1) cannot cross and so we have found data for
(4.1) with the property that the corresponding solutions break down in
finite time. Moreover these data can be arbitrarily close to homogeneity.

\vskip .5cm
{\bf Acknowledgment} We would like to thank
J\"urgen Ehlers and Bernd Schmidt for enlightening
discussions.
O.A.R. wants to thank the Max-Planck-Institut
f\"ur Astrophysik for hospitality and the DFG for
financial support.
\vskip .5cm
\vfill
\eject
\noindent
{\bf References}

\noindent
1. Barrow, J. D., G\"otz, G.: Newtonian no-hair theorems. Class.
Quantum Grav. {\bf 6} (1989) 1253-1265.
\next
2. Bildhauer, S., Buchert, T., Kasai, M.: Solutions in Newtonian
cosmology: the pancake theory with cosmological constant.
Astronomy and Astrophysics {\bf 263} (1992) 23-29
\next
3. Brauer, U.: An existence theorem for perturbed Newtonian
cosmological models. J. Math. Phys. {\bf 33} (1992) 1224-1232.
\next
4. Christodoulou, D., Klainerman, S.: The global nonlinear
stability of the Minkowski space. Princeton University Press (1993).
\next
5. Ehlers, J.: The Newtonian limit of general relativity. In
Ferrarese, G. (ed.) Classical mechanics and relativity: relationship
and consistency. Bibliopolis, Naples (1991).
\next
6. Ellis, G. F. R. in B. Bertotti, F. de Felice, A. Pascolini (eds.)
General relativity and gravitation. Reidel, Dordrecht (1984).
\next
7. Friedrich, H.: Existence and structure of past asymptotically
simple solutions of Einstein's field equations with positive
cosmological constant. J. Geom. Phys. {\bf 3} (1986) 101-117.
\next
8. Heckmann, O., Sch\"ucking, E.: in L. Witten, (ed.) Gravitation,
an introduction to current research. Wiley, New York (1962).
\next
9. Holm, D. D., Marsden, J. E., Ratiu, T., Weinstein, A.:
Nonlinear stability of fluid and plasma equilibria. Physics
Reports {\bf 123} (1985) 1-116.
\next
10. John, F.: Nonlinear wave equations, formation of singularities.
American Mathematical Society, Providence (1990).
\next
11. Klainerman, S., Majda, A.: Singular limits of quasilinear
hyperbolic systems with large parameters and the incompressible
limit of compressible fluids. Commun. Pure Appl. Math. {\bf 34}
(1981) 481-524.
\next
12. Kobayashi, S., Nomizu, K.: Foundations of differential geometry.
Interscience, New York (1963).
\next
13. Majda, A.: Compressible fluid flow and systems of conservation
laws in several space variables. Springer Verlag (1984).
\next
14. Makino, T.: On a local existence theorem for the evolution
equation of gaseous stars. In Nishida, T. (ed.) Patterns and
waves. North Holland, Amsterdam (1986).
\next
15. Peebles, P. J. E.: The large-scale structure of the universe.
Princeton University Press (1980).
\next
16. Rein, G., Rendall, A. D.: Global existence of classical solutions
to the Vlasov-Poisson system in a three dimensional cosmological
setting. Arch. Rat. Mech. Anal. (to appear)
\next
17. Sideris, T.: Formation of singularities in three-dimensional
compressible fluids. Commun. Math. Phys. {\bf 101} (1985) 475-485.
\next
18. Wald, R.: Asymptotic behaviour of homogeneous cosmological
models in the presence of a positive cosmological constant.
Phys. Rev. D {\bf 28} (1983) 2118-2120.

\end